\documentclass{article}

\usepackage{graphicx}

\def\1{\'{\i}}
\def\3{\ss}

\renewcommand{\thefootnote}{\fnsymbol{footnote}}

\def\bfalpha{\mbox{\boldmath $ \alpha $}}
\def\bfbeta{\mbox{\boldmath $ \beta $}}
\def\bfeta{\mbox{\boldmath $ \eta $}}

\def\bftheta{\mbox{\boldmath $ \theta $}}
\def\bfxi{\mbox{\boldmath $ \xi $}}

\title{A Computer Program to visualize Gravitational Lenses}

\author{Francisco ~Frutos Alfaro \\
       {\small Department of Physics and Space Research Center} \\ 
       {\small University of Costa Rica, San Jos\'e, Costa Rica} \\
       {\small {\tt frutos@fisica.ucr.ac.cr}}}

\date{\today}

\begin{document}

\thispagestyle{empty}

\maketitle

\renewcommand{\thefootnote}{\arabic{footnote}}

\pagestyle{headings}

\begin{abstract}
\noindent
Gravitational lenses are presently playing an important role in astrophysics. 
By means of these lenses the parameters of the deflector such as its mass, 
ellipticity, etc. and Hubble's constant can be determined. Using C, 
{\it Xforms}, {\it Mesa} and {\it Imlib} a computer program to visualize this 
lens effect has been developed. This program has been applied to generate 
sequences of images of a source object and its corresponding images. It has 
also been used to visually test different models of gravitational lenses.
\end{abstract}

\section{Introduction}

\noindent
Computer visualization is nowadays an important field of scientific research, 
because it often allows a better understanding of natural phenomena. 

\noindent
Gravitational lenses (hereafter GL), predicted by Einstein's general 
relativity theory, deflect the light rays coming from a distant object. GL's 
allow the formation of mutiple images of the same source object. Moreover 
this gra\-vi\-ta\-tional effect opens the possiblity of determining the mass 
of the deflector and the computation of the age of the Universe 
(Hubble's constant), as well as other lens parameters and cosmological 
parameters.

\noindent
An interactive computer program to visualize this gravitational effect has 
been developed. The program was written in the C programming language and uses 
the {\it Xforms} (tool kit), {\it Mesa} (graphic library) and {\it Imlib} 
(image library). The program runs on systems with Linux or Unix. An SGI 
version of this program is also available.

\noindent
In the first section the gravitational lens effect is briefly reviewed. The 
visualization program is presented in the second section. In the third section 
some applications are shown: image sequences and easy visual modelling. 
Improved versions of this program can be produced and the author wishes to 
invite the interested reader to participate in this process. The author is 
preparing a website for downloading this program:

\noindent
{\tt http://cinespa.ucr.ac.cr/software/xfgl/index.html}

\section{The Gravitational Lens Theory}

\subsection{The Gravitational Lens Equation}

\noindent
Due to the curvature of space-time, light rays coming from a distant 
source object (quasar, star or galaxy) are deflected when passing close to a 
lens or deflector (star, galaxy or cluster of galaxies). For weak 
gravitational fields the {\it Post-Newtonian} approximation is applicable. 
Under this approximation the deflection angle or Einstein's angle does not 
depend on the direction of propagation and the trayectory are approximated by 
straight lines. This gravitational light deflection is depicted in figure 1. 
From this figure an equation, the {\it GL equation} or the {\it ray tracing 
equation}\footnotemark[1] which is obeyed by light rays passing near a lens 
object can be deduced:

\begin{equation}
\label{ecgl} 
{\bf y} = {\bf x} - {{\bfalpha}}{(\bf x)} \; \; , 
\end{equation}

\noindent
where 

\begin{equation}
\label{ec16} 
{{\bfalpha}} ({\bf x}) = {{1} \over {\pi}} \int \! \kappa({\bf x} ')
{{({\bf x} - {\bf x} ')} \over {|{\bf x} - {\bf x} '|^2}} d^2 {\bf x} ' 
\; \; ,
\end{equation}

\begin{equation}
\label{ec17} 
{\bf y} = {{\bfeta} \over {\eta_{0}}} = {{D_{s} {\bfbeta}} \over {\eta_{0}}} 
\qquad , \qquad {\bf x} = {{\bfxi} \over {\xi_{0}}} 
= {{D_{l} {\bftheta}} \over {\xi_{0}}} \qquad , \qquad 
\eta_{0} = {{D_{s}} \over {D_{l}}} \xi_{0} 
\end{equation}

\noindent
and $ \xi_{0} $ is the length scale on the lens surface. The angle 
$ {{\bfalpha}} ({\bf x}) $ is called the Einstein's angle. The surface 
density of the lens is given by

\begin{equation}
\label{ec18} 
\kappa ({\bf x}) = {{\Sigma ({\bfxi})} \over {{\Sigma}_{cr}}} \; \; , 
\end{equation}

\noindent
where $ \Sigma_{cr} $ represents the {\it critical density}: 

\begin{equation}
\label{ec19}
\Sigma_{cr} = {{c^2 D_{s}} \over {4 \pi D_{l} D_{ls}}} \; \; . 
\end{equation}

\noindent
A generalization of equation (\ref{ecgl}) which considers the perturbation of 
a {\it ma\-cro\-lens} (a galaxy or cluster of galaxies) is given by

\begin{equation}
\label{ecgl2} 
{\bf y} = {\cal M} \cdot {\bf x} - {{\bfalpha}}{(\bf x)} \; \; , 
\end{equation}

\noindent
where the matrix $ {\cal M} $ is given by the expression

$$ {\cal M} = \left(\begin{array}{ccc}
1 - \kappa - \gamma \cos{\phi} & - \gamma \sin{\phi}   \\
- \gamma \sin{\phi}                & 1 - \kappa + \gamma \cos{\phi}
\end{array}\right) \; \; . $$

\noindent
The components of the matrix depend upon the parameters $ \kappa $ and 
$ \gamma $ and $ \phi $ which are respectively the dimensionless constant 
surface mass density, the dimensionless shear of the macrolens and the shear 
angle. The microlens (a star in a galaxy or a galaxy in a cluster of galaxies) 
is represented by means of the Einstein's angle $ {{\bfalpha}} ({\bf x}) $. 

\begin{figure}
\label{fig1}
\noindent
\centering
\includegraphics[width =\textwidth]{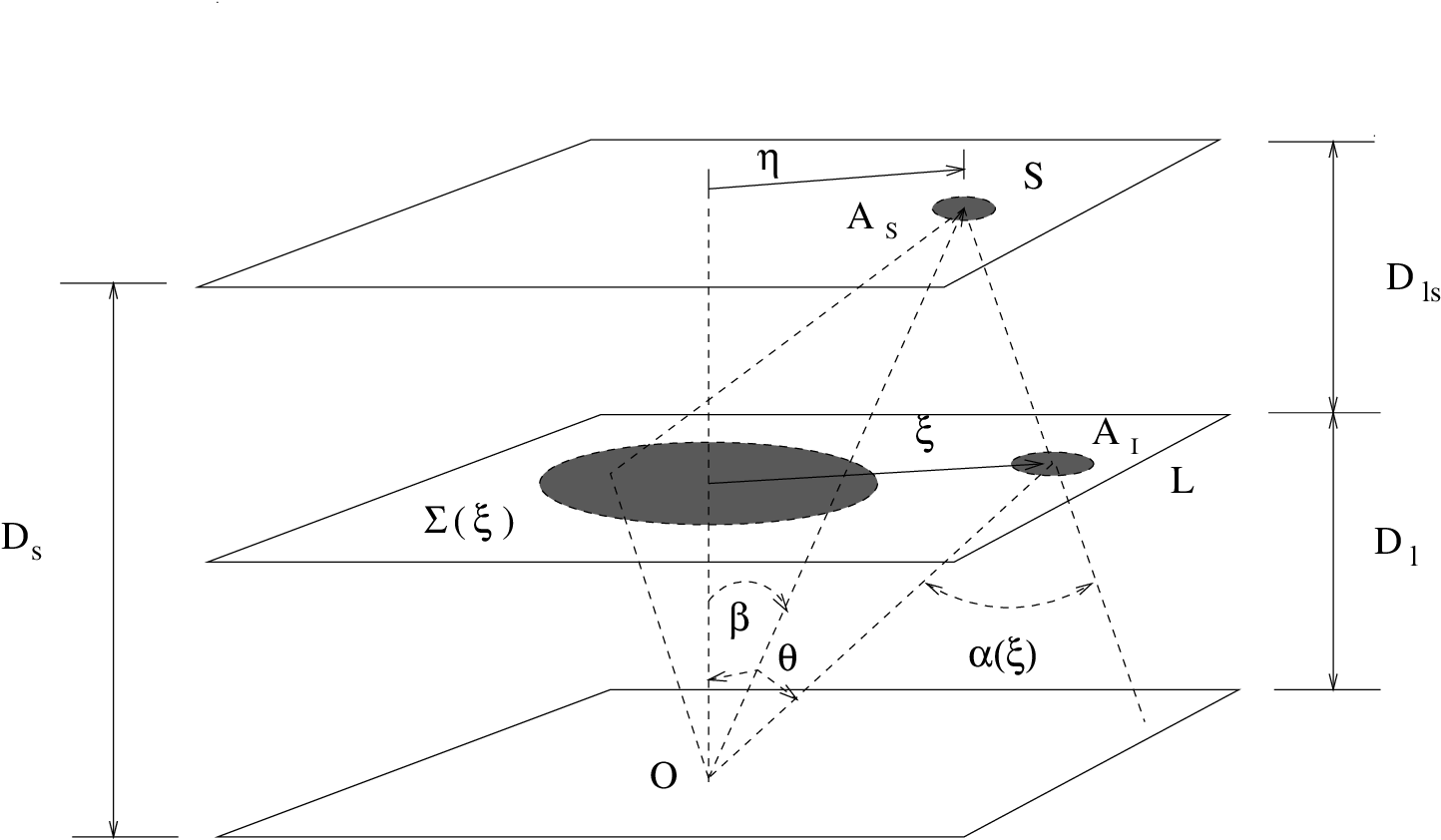}
\caption[Light Deflection]{The Post-Newtonian approximation of the light 
deflection}
\end{figure}

\subsection{Gravitational Lens Models}

\noindent
To simulate Einstein's angle there are several models for the lens mass 
distribution. The gravitational lens models are divided as follows:

\begin{itemize}      
\item Parametric models
\item Non-parametric models.
\end{itemize}

\noindent
At the present time the program has only parametric 
models\footnotemark[1]$ ^{-} $\footnotemark[3]. The non-parametric models are 
lately used to model gravitational 
lenses\footnotemark[4]$ ^{-} $\footnotemark[6]. For weak lensing the 
{\it Kaiser-Squires} method can be applied to model lenses
\footnotemark[7]$ ^{-} $\footnotemark[10]. In the program these techniques 
have not been implemented. 

\noindent
The program includes the following parametric models:

\begin{itemize}
\item Chang Refsdal Model,
\item Double Plane Lens,
\item Transparent Sphere,
\item Singular Isothermal Sphere,
\item Nonsingular Isothermal Sphere,
\item Elliptical Model,
\item King Model,
\item Truncated King Model,
\item Hubble Model,
\item De Vaucouleur Model,
\item Spiral Model,
\item Multipole Lens,
\item Rotation Lens and
\item Uniform Ring.
\end{itemize}

\noindent
The interested reader is refered to the literature for a discussion of these 
models\footnotemark[1]$ ^{-} $\footnotemark[3].

\section{The Visualization program}

\subsection{Description of the program}

\noindent
The program was written in C and the {\it Mesa graphic libraries} (free 
version of the Open GL) have been used. These graphic subroutines are 
available for Unix or Linux systems. The {\it XForms Library} was used to 
designed the control panel program\footnotemark[12] (see fig. 2). The 
{\it Image Library} permits to load an image file on the program. These 
libraries can be found at the following addresses: \\
\noindent 
{\tt http://bragg.phys.uwm.edu/xform} \\
\noindent 
{\tt http://www.mesa3d.org/} \\
\noindent
{\tt ftp://ftp.enlightenment.org/pub/enlightenment/imlib/}

\noindent
The program creates a window: the {\it Control Panel} (see fig. 2). The user 
can control all item on it by clicking. A second window, the 
{\it Image window}, appears when the Image Window button is clicked on this 
panel (see fig. 3). The Help button on the panel give the user a concise 
program guide. 

\noindent
A version of this program which employs the SGI 
{\it Graphic Libraries}\footnotemark[11] is also available.

\begin{figure}[p]
\label{general}
\centering
\includegraphics[width =\textwidth]{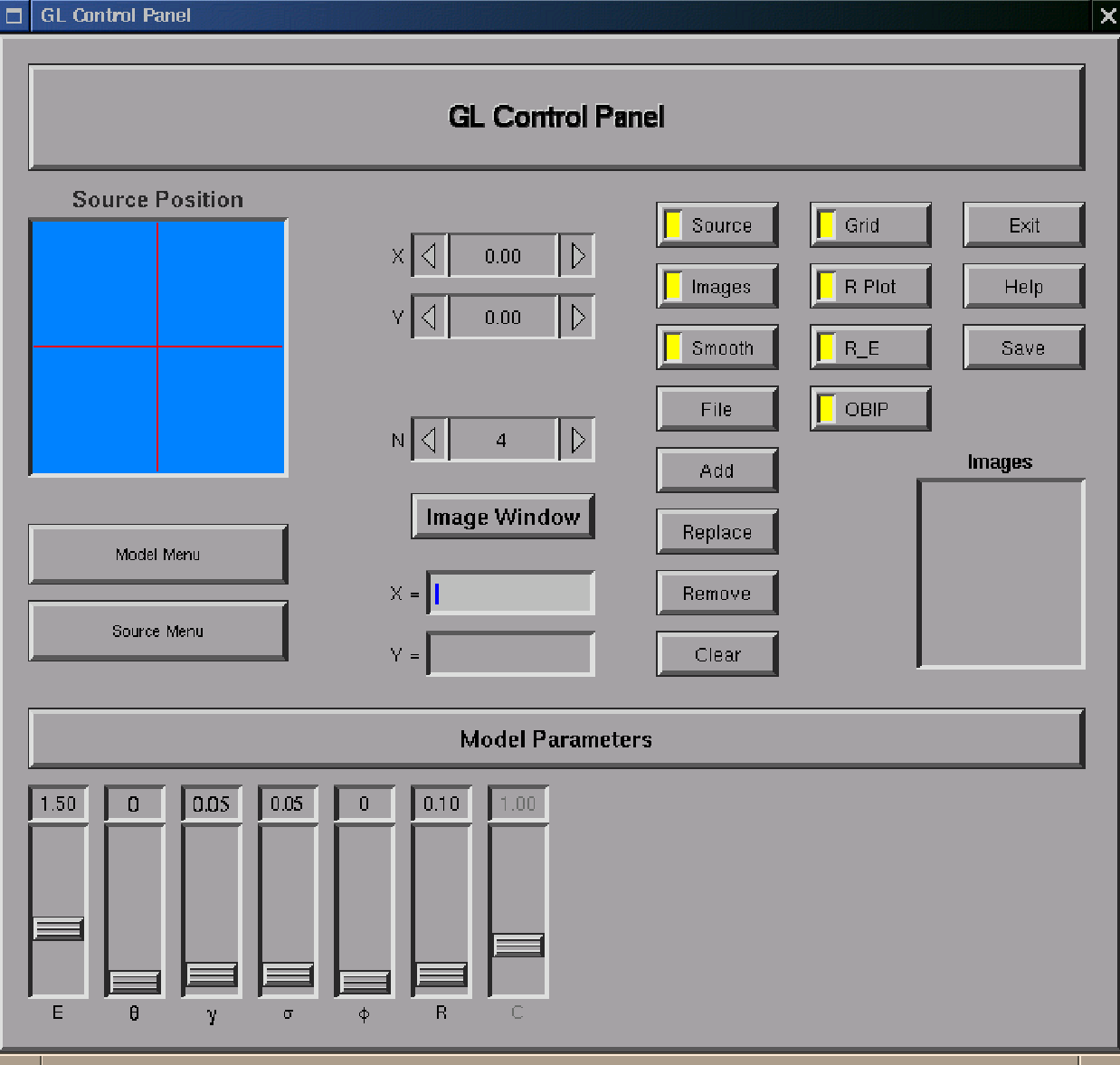}
\caption[Control Panel]{Control Panel}
\end{figure}

\begin{figure}[p]
\label{images2}
\centering
\includegraphics[width =\textwidth]{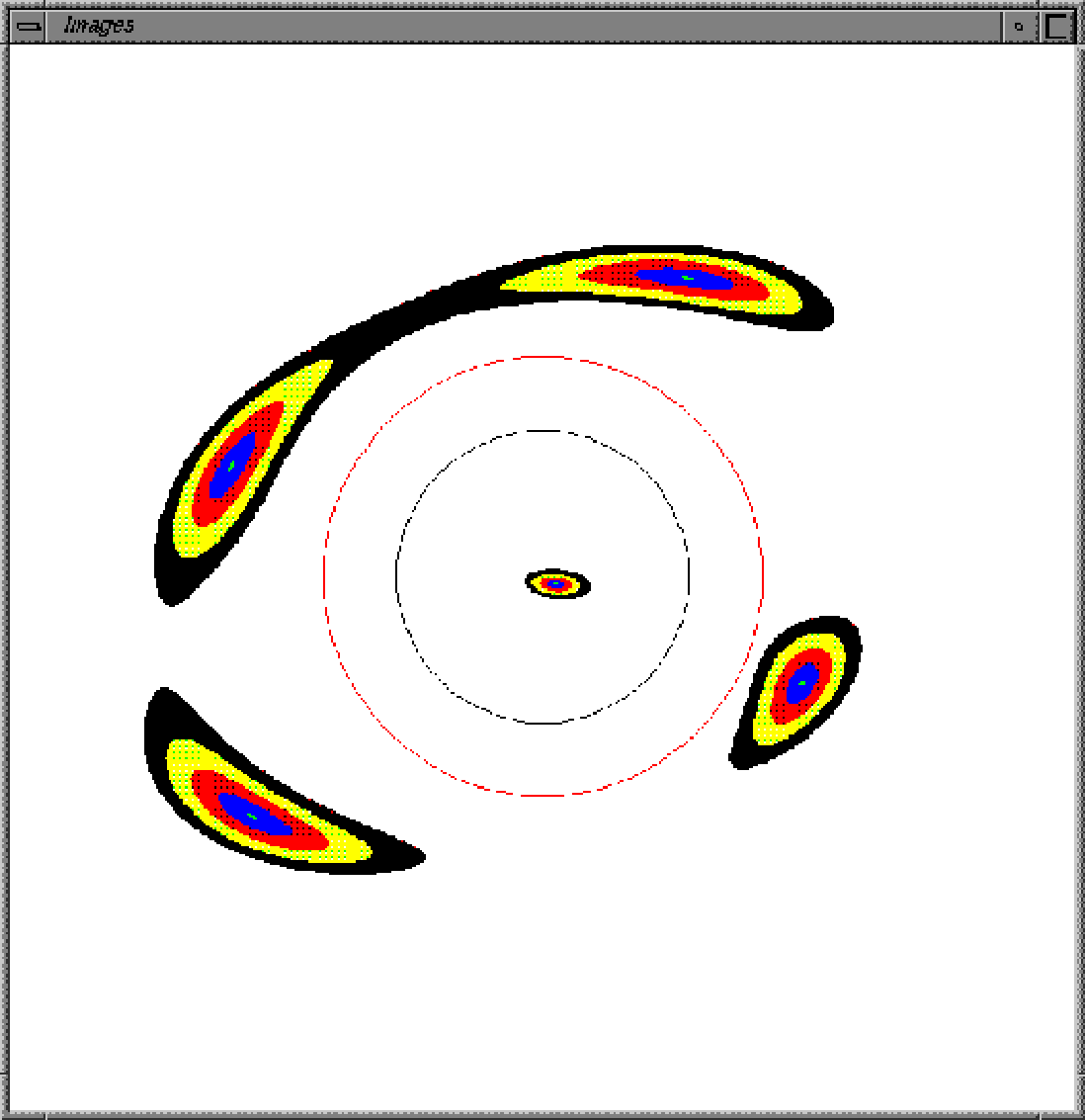}
\caption[The Image Window]{The Image Window}
\end{figure}

\subsection{The Control Panel}

\noindent
All the objects are adjusted interactively with the mouse. This control panel 
interface has the following items:

\begin{itemize}
\item Source Menu (Filled Circle, Coloured Rings and Image File)
\item Model Menu (the aforementioned models)
\item Source Positioner
\item Three counters, two for source positioning (X, Y) and one for adjusting 
the pixel resolution (N) 
\item Two inputs for positioning the observed lense images (X, Y)
\item Buttons (Model Paramters, Image Window, Source, Images, Smooth, Grid, 
R Plot, R\_E, File, Add, Replace, Remove, Clear, OBIP, Save, Help, 
Exit), 
\item Sliders (if one chooses a model, then the corresponding sliders appear) 
\item A browser for showing the observed lense images
\end{itemize}

\noindent
For a detailed discussion of these items see the manual program (to get the 
program manual send an email to the author or click the Help button on the 
control panel).

\subsection{The Image Window}

\noindent
On this window the events (images, ray plot, etc.) appear. All variations of 
the parameters on the control panel are shown immediately on this window. To 
show the images of a chosen gravitational lens model the {\it ray tracing 
method} has been used. The caustic is re\-pre\-sented by means of the 
{\it ray plot method}\footnotemark[1]. 

\section{Applications}

\noindent
In this section two of many applications of the program are discussed:

\begin{itemize}
\item {\it Image sequences} 
\item {\it Easy visual modeling} 
\end{itemize}

\noindent
An image sequence for an elliptical model and a visual modeling of the 
gra\-vi\-ta\-tional lens $ 2237 + 0305 $ will be shown.

\subsection{Image sequences}

\noindent
An image sequence is easy to generate with this program. The structure of the 
images for different positions of the source can be investigated by means of 
such an image sequence.

\bigskip
\noindent
{\sl An Elliptical Lens}
\smallskip

\noindent
A sequence of images is shown in figure 4. The sequence begins on the top 
left-hand-side panel of the sequence. The source, concentric rings, 
moves from left to right. The source is not shown, only the ima\-ges. The 
best possible resolution given by the control panel was used. In this sequence 
the shear $\gamma $ and the constant surface mass density $ \sigma $ are non 
zero and the shear angle $ \phi = 0 $. There are two circles: the Einstein 
ring or lens scale and the core radius. The formation of a small and a large 
arc can be seen. In the middle of the sequence one can recognize the 
deformation of the Einstein ring in an ellipse. Due to the core radius an 
image appears in the centre, which fuses with an arc, in the case that the
source moves away.

\begin{figure}[p]
\label{ellipt}
\centering
\includegraphics[width =\textwidth]{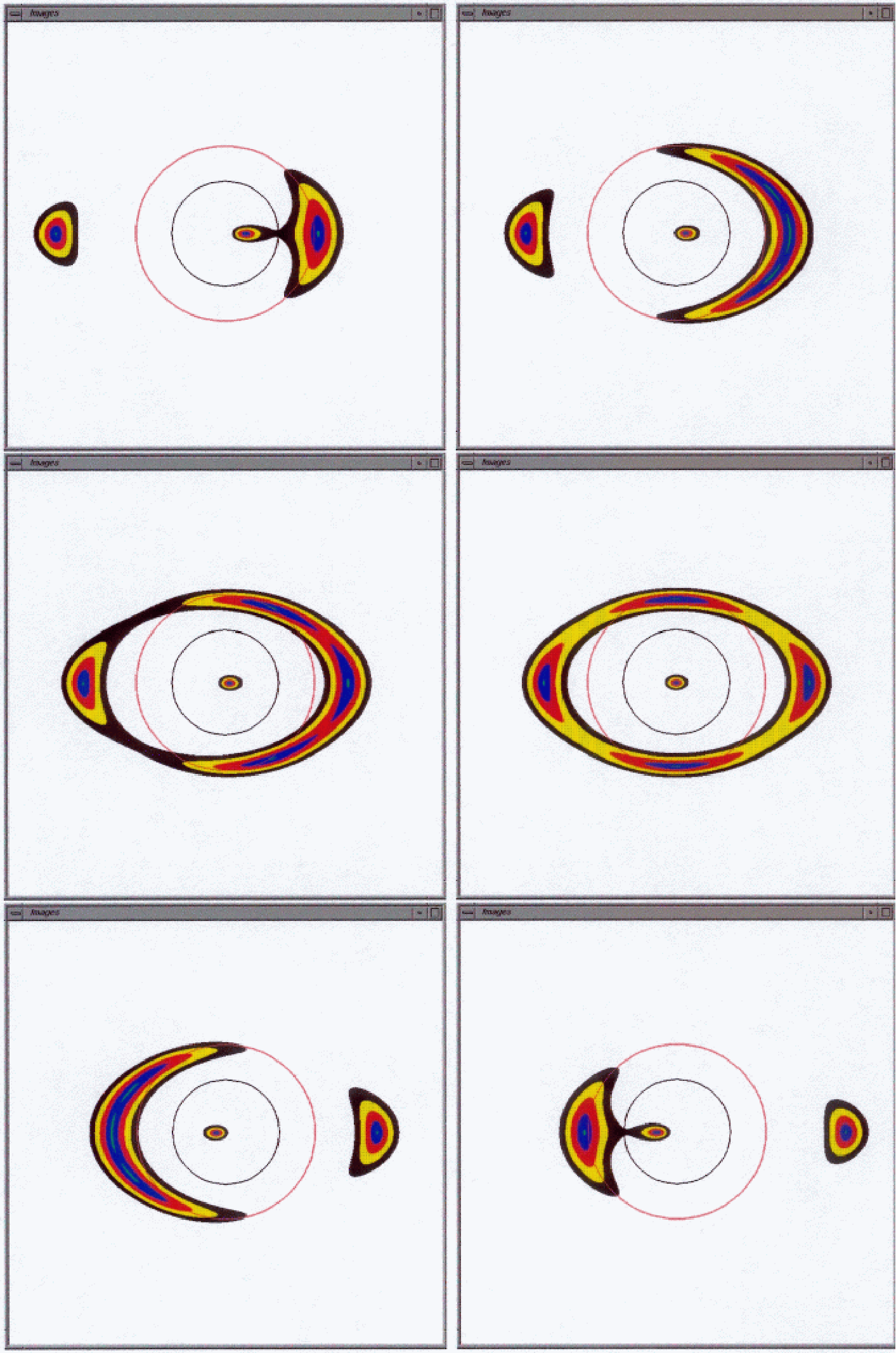}
\caption[An elliptical lens]{An elliptical lens}
\end{figure}

\begin{figure}[p]
\label{GL2237}
\centering
\includegraphics[width =\textwidth]{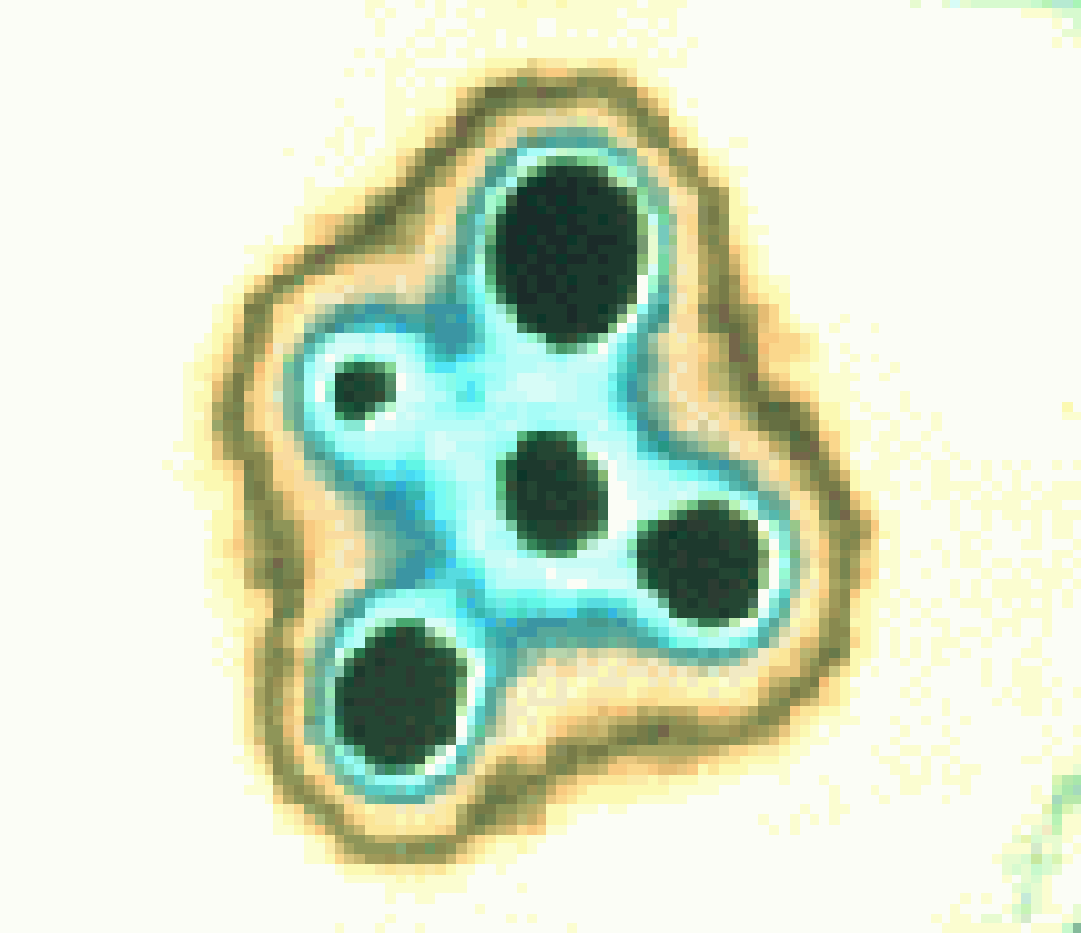}
\caption[GL $ 2237 + 0305 $]{Gravitational lens $ 2237 + 0305 $}
\end{figure}

\begin{figure}[p]
\label{m2237el}
\centering
\includegraphics[width =\textwidth]{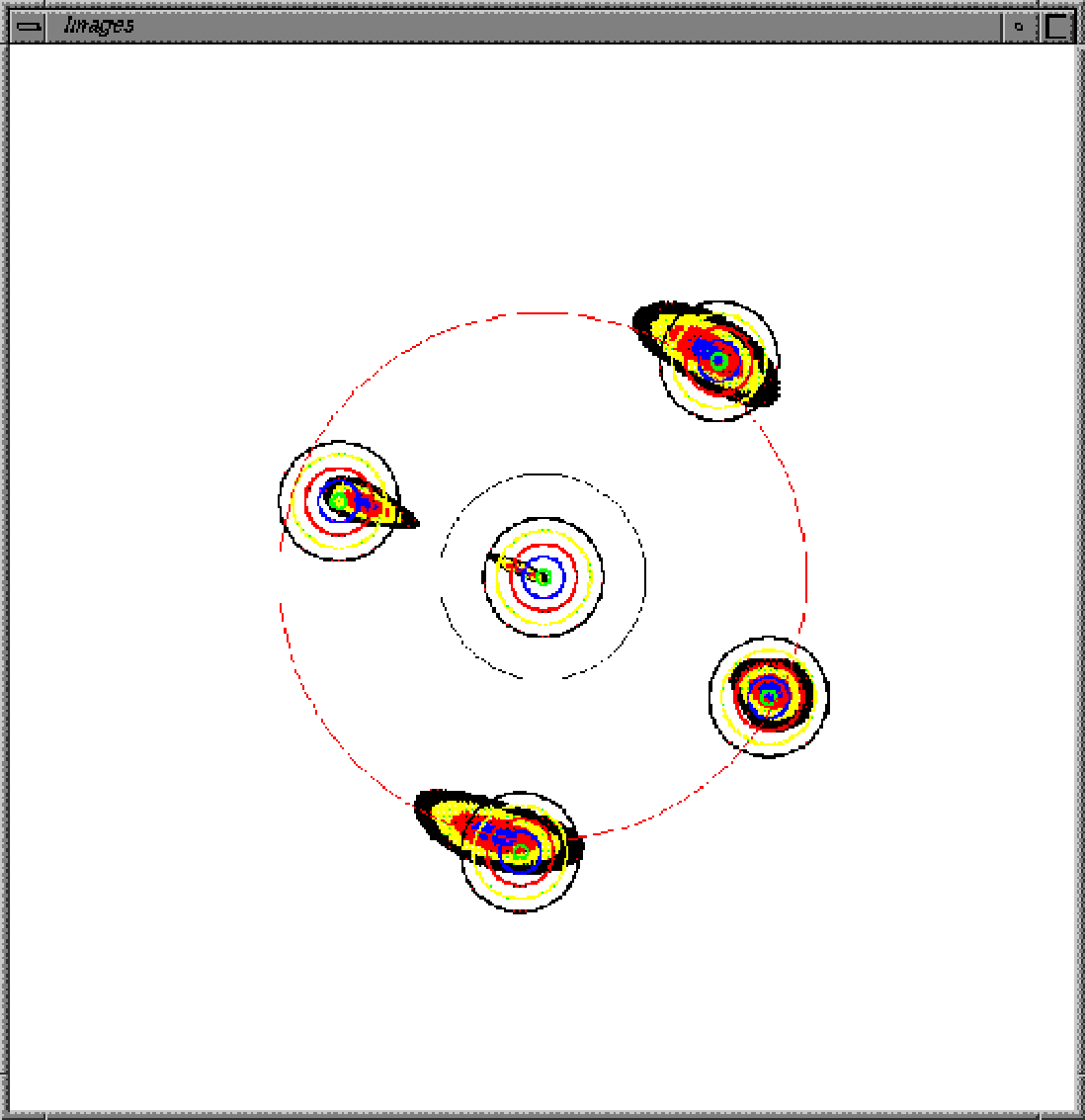}
\caption[Visual fitting of the system $ 2237 + 0305 $]{Visual fitting of the 
gravitational lens $ 2237 + 0305 $}
\end{figure}

\subsection{Easy Visual Modeling}

\noindent
Here an easy visual modeling of the gravitational lens system 
$ 2237 + 0305 $ is shown. The model images are considered well-fit when they 
overlap most of the area of the observed images 
(represented through concentric circles), as visually estimated. At the 
present time a fitting subroutine is not implemented. When such a subroutine 
is included, this program would become an excellent tool to model 
gravitational lenses.

\bigskip
\noindent
{\sl Visual Model for the Gravitational Lens $ 2237 + 0305 $}
\smallskip

\noindent
The system $ 2237 + 0305 $ [see fig. 5] was discovered in 
1985\footnotemark[13]. Due to its unusual shape this lens is called 
{\it Einstein cross}. Attempts to model this lens have been carried 
out\footnotemark[14]$ ^{-} $\footnotemark[20]

\noindent
The data used in tables \ref{tab1} and \ref{tab2}, was taken from Kent and 
Falco, and Schneider et al.\footnotemark[15]$ ^{-} $\footnotemark[16] The 
cosmological distances of table \ref{tab2} were calculated using the equations 
from Schneider et al.\footnotemark[15] 
($ H_{0} = 100 \, {\rm km} \, {\rm s}^{- 1} \, {\rm Mpc}^{- 1} $ and 
$ q_{0} = 1 / 2 $). 

\begin{table}[ht]
\caption{Image positions for the lens $ 2237 + 0305 $}
\smallskip
\label{tab1}
\begin{center}
\begin{tabular}{c c c}
\hline
                 & \multicolumn{2}{c}{Positions}   \\
 Object         &        X        &       Y       \\
                &    $ ('') $     &   $ ('') $    \\
\hline
 2237 $ + $ 0305 A                & $ 0.08 \pm 0.01 $  
                & $ - 0.94 \pm 0.01 $           \\
 2237 $ + $ 0305 B                & $ - 0.60 \pm 0.02 $  
                & $ 0.74 \pm 0.01 $             \\
 2237 $ + $ 0305 C                & $ 0.70 \pm 0.01 $  
                & $ 0.26 \pm 0.01 $             \\
 2237 $ + $ 0305 D                & $ - 0.77 \pm 0.02 $   
                & $ - 0.41 \pm 0.02 $           \\
 Galaxy          & $ 0.00 \pm 0.02 $   
                 & $ 0.00 \pm 0.02 $             \\
\hline 
\end{tabular}
\end{center}
\end{table}

\begin{table}[ht]
\caption{Parameters of the lens galaxy and the source}
\smallskip
\label{tab2}
\begin{center}
\begin{tabular}{l c}
\hline
 Parameter      &             Value               \\
\hline
 $ z_{l} $      &          $ 0.0394 $            \\
 Distance $ D_{l} $             &  $ 110 \, {\rm Mpc} $ \\
 Ellipticity $ \epsilon $       &  $ 0.57 $      \\
 Angle of the mayor axis        &  $ 77^{\circ} $ \\
 Angle of the bulk     &  $ 38^{\circ} $          \\
 Velocity dispersion $ \sigma_v $ & $ 215 \, {\rm km} / {\rm s} $ \\
  $ z_{s} $      &          $ 1.695 $             \\
Distance $ D_{s} $  &      $ 867 \, {\rm Mpc} $   \\
 Distance $ D_{ls} $ &      $ 825 \, {\rm Mpc} $   \\
\hline
\end{tabular}
\end{center}
\end{table}

\noindent
Since the positions of the observed images [table \ref{tab1}] are too small 
to be shown in the image window, a scale factor of $ 200 $ was chosen 
(the scale factor can be chosen by the user). Then all positions are 
multiplied by this scale. The positions are also rotated around the $ X $ axis.

\noindent
The lens galaxy is elliptical, so it is reasonable to choose an elliptical 
model in the program, although one can also try other 
models\footnotemark[15]$ ^{-} $\footnotemark[18]. A lens model of $ N $
($ N = 6724 $) point masses was used by Schenider et al.\footnotemark[15].
An {\it ellip\-tical King} model and an {\it ellip\-tical de Vau\-cou\-leur}
model were used in Kent and Falco\footnotemark[16]. In Rix et al. a the
{\it de Vau\-cou\-leur} model was used\footnotemark[17]. A general SIS model
with shear was used in Wambsganss and Paczynski\footnotemark[18].

\noindent
The lens major axis has a particular orientation, therefore the coordinate 
system is rotated around an angle $ \theta $, so that the elliptical potential 
has the same orientation. The angle $ \theta $ is taken as a parameter. The 
constant surface mass density and the shear are zero. The visual fitting is 
shown in fig. 5. In table \ref{tab3} the model parameters of the elliptical 
model are shown. These were found through variation of sliders and source 
positioners.

\begin{table}[ht]
\caption{Model parameter of the elliptical model and source position}
\smallskip
\label{tab3}
\begin{center}
\begin{tabular}{l c}
\hline
 Parameter      &             Value                    \\
\hline
 E: Lens scale  &          $ 180 $                     \\
 $ \theta $: Rotation angle & $ 336^{\circ} $          \\
 $ \gamma $: Shear      &  $ 0 $                       \\
 $ \sigma $: Constant surface mass density   & $ 0 $   \\
 $ \phi $: Shear angle  & $ 0^{\circ} $                \\
 R: Source Radius       &  $ 50 $                      \\
 C: Core scale  &          $ 70 $                      \\
 $ \epsilon $: Ellipticity    & $ 0.54 $              \\
 $ \alpha $: Softness  & $ 0.05 $                     \\
 $ \kappa $: Constant Central mass density & $ 1.00 $ \\
 (X, Y): Source Position & $ (28.0, \, 12.0) $       \\
\hline
\end{tabular}
\end{center}
\end{table}

\noindent
The data for the lens scale ({\it Einstein Radius}), the core scale and the 
source position must be divided by $ 200 $. So one obtained 
$ {\alpha}_{E} = 0''.90 $ , $ \theta_{c} = 0''.35 $ and 
$ (X, \, Y) = (0''.14, \, 0''.06) $ . The Einstein radius and core radius are 
determined by means of the equations $ {\cal R}_E = {\alpha}_{E} D_{l} $ and 
$ r_{c} = \theta_{c} D_{l} $ ($ {\cal R}_E = 0.48 \, {\rm Kpc} $ and 
$ r_{c} = 0.19 \, {\rm Kpc} $). 

\noindent
From the following equation the lens mass can be found 

$$ {\cal R}^2_E = \frac{4 G M}{c^2} \frac{D_{l} D_{ls}}{D_{s}} \; \; . $$

\noindent
The velocity dispersion is determined through an equation from Blandford and 
Kochanek\footnotemark[21]$ ^{-} $\footnotemark[22].

\noindent
The major axis angle can be calculated from $ \theta' = 90^{\circ} + \theta $ 
($ \theta = - 24^{\circ} $). 

\noindent
In table \ref{tab4} all values for the different mo\-dels are shown: the 
elliptical King model (model A)\footnotemark[16], the de Vaucouleur model 
(mo\-del B, the source position is referred to the image A)\footnotemark[17], 
the generalized SIS model (model C)\footnotemark[18] and the elliptical model. 
The models A, B and C were fitted. Our elliptical model is not reliable, 
because the program does not possess a fitting subroutine, but one can 
estimate and compare with other models.
 
\begin{table}[ht]
\caption{Model parameter for the system 2237 $ + $ 0305}
\smallskip
\label{tab4}
\begin{center}
\begin{tabular}{l c c c c}
\hline
                 & \multicolumn{4}{c}{Model}       \\
 Parameter       & A   &  B  &  C  & Elliptical    \\
\hline
 Einstein Ring  & $ 0''. 90 $ & $ 0''. 90 $ & $ 0''. 874 $ & $ 0''. 90 $ \\
 Shear          & $ - $ & $ - $   & $ 0.02 $     &  $ - $               \\
 Angle of the mayor axis $ ^{\circ} $ 
                & $ 66. 8 $ & $ 68 $ & $ 66. 84 $ & $ 66 $          \\
 Core Radius kpc & $ 0. 10 $ & $ - $ & $ - $       & $ 0. 19 $   \\
 Ellipticity $ \epsilon $ & $ 0. 42 $ & $ 0. 3 $ & $ - $ & $ 0. 54 $ \\
 Softness $ \alpha $ & $ - $ & $ - $ & $ - $ &   $ 0. 05 $             \\
 Central surface mass density $ \kappa $ 
                & $ - $ & $ - $ & $ - $ & $ 1. 00 $ \\
 Velocity  dispersion km/s & $ 170 $ & $ 209 $    & $ - $ & $ 109 $    \\
 Lens Mass $ 10^{10} {\cal M}_{\odot} $
                & $ 1. 2 $ & $ 1. 08 $ & $ 1. 49 $       & $ 1. 15 $     \\
 Source Position &    &     &     &               \\
 X              & $ - 0''. 07 $ & $ 0''. 159 $    & $ - $ & $ 0''. 14 $  \\
 Y              & $ - 0''. 02 $ & $ 0''. 877 $ & $ - $ & $ 0''. 06 $     \\
\hline
\end{tabular}
\end{center}
\end{table}

\section{Conclusions and Future Work}

\noindent
From both the didactical and scientific point of view a program to visualize 
the gravitational lenses is useful. This versatile program works quickly and
interactively with the mouse. 

\noindent
With this computer program the user has a tool to visualize and to visually 
model gravitational lenses. The applications of the program, we have showed in 
this paper, are:

\begin{itemize}
\item {\it Sequences of images}
\item {\it Easy visual modeling}
\end{itemize}

\noindent
The user can produce sequences of images for a chosen gra\-vi\-tational lens 
model. Through the variation of model pa\-ra\-meters he or she can investigate 
the structure of the images. The user can also attempt to visually model 
observed gravitational lenses. The observed position data can be used as
input data and the model parameters can be easily varied in order to
approximate the observed images. So the user can quickly obtain model
parameter estimations. Some observed lenses have already been mode\-led and
the user can compare those results with the output from a chosen model of the
control panel.

\bigskip
\bigskip

\noindent
{\sl Future Work}
\smallskip

\noindent
The program can be improved by the inclusion of some additional subroutines:

\begin{itemize}
\item {\it Contour subroutine for the isochrones (time delay)}
\item {\it Light curves subroutine (dependence of brightness with time)} 
\item {\it Subroutine for computing the image magnification}
\item {\it Subroutine to calculate critical curves and caustics}
\item {\it Fitting subroutine}
\item {\it Root finder subroutine}
\item {\it Subroutine to load images of observed gravitational lenses}
\item {\it Subroutine with more complex (elliptical) models}
\item {\it Subroutine for superposition of models in different lens planes}
\item {\it Subroutine with cosmic string lens models}
\item {\it Subroutine for non-parametric reconstruction} 
\item {\it Kaiser-Squires Subroutine} 
\end{itemize}

\noindent
The author is working on implementing some of the abovementioned improvements. 

\bigskip
\bigskip
\noindent
{\it Acknowledgments:} The author would like to thank M. Magall\'on for his 
contribution in developing part of the software discussed in this paper.

\bigskip
\bigskip
\bigskip

\newpage

\noindent
{\bf References}
\bigskip

\noindent
\footnotemark[1]\footnotesize{P. Schneider, J. Ehlers, E.E. Falco, 
{\sl Gravitational Lenses}, A\&A Library, Springer, Heidelberg, 1992.}

\noindent
\footnotemark[2]\footnotesize{J. Huchra, 
{\sl Galatic Structure And Evolution}, in R. Meyers, 
{\sl Encyclopedia of Astronomy and Astrophysics}, 203-19, Academic Press, 
Inc., London, 1989.}

\noindent
\footnotemark[3]\footnotesize{F. Frutos-Alfaro, {\sl Die interaktive 
Visualisierung von Gra\-vi\-ta\-tions\-lin\-sen}, PHD Thesis, 
Eberhard-Karls-Universit\"at T\"ubingen, 1998.}

\noindent
\footnotemark[4]\footnotesize{P. Saha, L. L. R. Williams, {\sl Non-parametric 
reconstruction of the galaxy-lens in PG 1115 + 080}, {\it astro-ph/9707356}, 
31 July 1997.} 

\noindent
\footnotemark[5]\footnotesize{H. M. Abdel Salam, P. Saha, L. L. R. Williams, 
{\sl Non-parametric reconstruction of cluster mass distribution from strong 
lensing: modelling Abell 370}, MNRAS, 294, 734-746, 1998.}

\noindent
\footnotemark[6]\footnotesize{H. M. Abdel Salam, P. Saha, L. L. R. Williams, 
{\sl Non-parametric Reconstruction of Abell 2218 from Combine Weak and Strong 
Lensing}, AJ, 116, 4, 1541-1552, 1998.}

\noindent
\footnotemark[7]\footnotesize{N. Kaiser, G. Squires, 
{\sl Mapping the Dark Matter with Weak Gravitational Lensing}, ApJ, 404, 
441-450, 1993.}

\noindent
\footnotemark[8]\footnotesize{P. Schneider, C. Seitz, 
{\sl Steps towards nonlinear cluster inversion gravitational distortions I}, 
A\&A, {\bf 294}, 411-431, 1995.}

\noindent
\footnotemark[9]\footnotesize{C. Seitz, P. Schneider, 
{\sl Steps towards nonlinear cluster inversion gravitational distortions II}, 
A\&A, {\bf 297}, 287-299, 1995.}

\noindent
\footnotemark[10]\footnotesize{C. Seitz, P. Schneider, 
{\sl Steps towards nonlinear cluster inversion gravitational distortions III}, 
A\&A, {\bf 318}, 687-699, 1997.}

\noindent
\footnotemark[11]\footnotesize{Silicon Graphics, 
{\sl Gra\-phics Li\-bra\-ry Pro\-gramming Gui\-de}, Si\-li\-con Gra\-phics, 
Inc., 1991.}

\noindent
\footnotemark[12]\footnotesize{M.H. Overmars, {\sl Forms Library: A Graphical 
User Interface Toolkit for Silicon Graphics Workstations}, Department of 
Computer Science, Utrecht University, 1991.}

\noindent
\footnotemark[13]\footnotesize{J. Huchra, M. Gorenstein, S. Kent, I. Shapiro, 
G. Smith, E. Horine, R. Perley, {\sl $ 2237 + 0305 $: A New and Unusual 
Gra\-vi\-tational Lens}, AJ, {\bf 90}, 691-6, 1985.}

\noindent
\footnotemark[14]\footnotesize{H. Yee, {\sl High-Resolution Imaging of the 
Gravitational Lens System Candidate $ 2237 + 030 $}, AJ, {\bf 95}, 1331-39, 
1988.}

\noindent
\footnotemark[15]\footnotesize{D. Schneider,  E. Turner, J. Gunn, J. Hewitt, 
M. Schmidt and C. Lawrence, {\sl High-Resolution CCD Imaging And Derived 
Gravitational Lens Models of $ 2237 + 0305 $}, AJ, {\bf 95}, 6, 1619-28, 1988.}

\noindent
\footnotemark[16]\footnotesize{S. Kent, E. Falco, {\sl A Model For The 
Gravitational Lens System $ 2237 + 0305 $}, AJ, {\bf 96}, 1570-74, 1988.}

\noindent
\footnotemark[17]\footnotesize{H. Rix, D. Schneider and J. Bahcall, 
{\sl Hubble Space Telescope Wide Field Camera Ima\-ging of the Gravitational 
Lens 2237 + 0305}, AJ, {\bf 104}, 959-67, 1992.}

\noindent
\footnotemark[18]\footnotesize{J. Wambsganss, B. Paczynski {\sl Parameter 
Degeneracy in Models of the Quadruple Lens System Q $ 2237 + 0305 $}, AJ, 
{\bf 108}, 1156-62, 1994.}

\noindent
\footnotemark[19]\footnotesize{M. Irwin, {\sl Photometric Variations in the 
Q 2237 + 0305 System: First Detection of a Microlensing Event}, A. J., 
{\bf 98}, 1989-94, 1989.}

\noindent
\footnotemark[20]\footnotesize{K-H. Chae, D.A. Turnshek, V.K. Khersonsky, 
{\sl Realistic Grid of Models for the Gra\-vi\-ta\-tionally Lensed Einstein 
Cross (Q2237+0305) and its Relation to Observational Constraints}, 
Ap. J., 495, 609-616, 1998.}

\noindent
\footnotemark[21]\footnotesize{R. Blandford, C. Kochanek, {\sl Gravitational 
Imaging By Isolated Elliptical Potential Wells. I. Cross Sections}, ApJ, 
{\bf 321}, 658-75, 1987.}

\noindent
\footnotemark[22]\footnotesize{C. Kochanek, R. Blandford, 
{\sl Gravitational Imaging By Isolated Elliptical Potential Wells. II. 
Probability Distributions}, ApJ, {\bf 321}, 676-85, 1987.}

\end{document}